\documentstyle[11pt,newpasp,twoside,epsf]{article}
\def\msun{M_{\odot}}
\def\gtorder{\mathrel{\raise.3ex\hbox{$>$}\mkern-14mu
             \lower0.6ex\hbox{$\sim$}}}
\def\ltorder{\mathrel{\raise.3ex\hbox{$<$}\mkern-14mu
             \lower0.6ex\hbox{$\sim$}}}

\def\edcomment#1{\iffalse\marginpar{\raggedright\sl#1\/}\else\relax\fi}
\reversemarginpar

\begin{document}

\title{Black Holes and Pulsar Binaries
}

\author{Steinn Sigurdsson}
\affil{525 Davey Laboratory, Department of Astronomy \& Astrophysics,
    Pennsylvania State University, Pa 16802}

\begin{abstract}
Stellar mass black holes are formed in the field, and are observed in binaries.
Population synthesis estimates for the formation of BH-PSR binaries
suggest these may be observed, but with very large formal uncertainties in
the formation rate.
In globular clusters, exchanges and binary interactions present more channels for BH-PSR
formation, while ejection through recoil can remove a large fraction of the initial
black hole population. 
A BH-PSR system ought to be observed in the near future, most likely in
a globular cluster.
\end{abstract}

%
{\bf 1. The Field}\\
We have observed several stellar mass black hole binary candidates in the field
(cf Bailyn et al 1998). Current best estimates are that the minimum ZAMS for black hole
formation is of the order $M_0 \sim 25 \msun$, and that the cut-off mass may have significant
metallicity dependence, with higher cut-off and narrower range of masses for black hole
formation at higher metallicities (Fryer \& Kalogera 2001, Fryer et al 2002).  
Possibly no stars, or only a very small proportion of stars, form black holes at
super-solar metallicities.  It is also possible that Pop III stars formed intermediate
mass ($M_{BH} \gtorder 100 \msun $) black holes directly in significant numbers
(Abel, Bryan, \& Norman 2002, Bromm, Coppi, \& Larson 1999). 
We infer that the fraction of black hole progenitor
systems in binaries is comparable to that of other field stars, and a significant
fraction of black holes remain bound to their companion after formation.
In the case of a neutron star companion, the natal kick on formation reduces
the probability of a bound system, but in general, for plausible amplitude kicks,
a finite fraction of systems remains bound. This is also the case if black holes
receive natal kicks in some fraction of instances (Brandt, Podsiadlowski, \& Sigurdsson 1995,
Israelian et al 1999, Kalogera et al 2001).
One approach to estimating the formation of BH-PSR binaries is through stellar population
synthesis models, which can either be done {\it ab initio}, or calibrated to observations.
We recently carried out such a synthesis (Sipior \& Sigurdsson 2002), as have a number
of other authors (cf Belczynski, K., Kalogera, V., \& Bulik 2002, Grishchuk et al 2001), 
using a range of independent stellar evolution codes
and model assumptions for the kinematics of the supernova events.

In constructing such a model, assumptions need to be made about: 

\noindent$\bullet$ mass function, $x_{IMF}$; binary fraction $f_{bin}$;\\
\noindent$\bullet$ cut-off mass for BH formation, $M_{0}$; BH mass function $P(M_{BH})$;\\
\noindent$\bullet$ mass ratio, $P(q)$; semi-major axis and eccentricity distribution, $P(a_i), P(e)$;\\
\noindent$\bullet$ natal kick distribution, $P(\Delta v)$; mass loss and mass transfer.\\
The net results of our models were in the median range for predicted NS-NS binaries, consistent
with the observed population, and in the median of the range explored by Kalogera et al's models.

For field BH-PSR binaries there are two primary channels:
\begin{itemize}
\item{I} BH+PSR - pulsar formed second, lifetime $\tau \ltorder 10^7$ years, the major question
is whether any such systems are currently being formed in the galaxy due to high metallicity of
the progenitor systems, and corresponding high mass-loss rates. Since star forming regions have
a range of metallicities in practise, we expect some such systems to be formed, but likely with
reduced efficiency compared to the past.
\item{II} BH+MSP - for every few thousand BH binaries formed, one may form with the neutron star
forming first, and remaining bound in tight enough an orbit for mass transfer leading
to a recycled, long lived pulsar, which then remains bound to the black hole after formation.
Since such systems are typically long lived, observable for $\tau \gtorder 10^9$ years,
and may have formed more efficiently in the past due to lower metallicities of stars formed then,
we predict the type II systems are actually more likely to be observed. 

The channel for formation provides a  prediction of the distribution of semi-major axis and 
eccentricity (Sipior \& Sigurdsson in preparation) and some correlation between the binary
angular momentum and pulsar spin, which may be observable. Such systems should be observable
at the rate of $O(10^{-4})$, ie we expect to see one in 5,000-10,000 PSRs as a BH-PSR binary.
So one should be observed in the near future.
\end{itemize}
{\bf 2. Clusters}\\
The existence of stellar mass black holes in the field implies the existence of stellar
mass black holes in globular clusters. Number fractions of black holes expected from
normal IMFs are of the order few$\times 10^{-4}$, and thus tens-to-hundreds of black
holes formed per cluster. With the lower metallicity of globulars
we expect, if anything, for stellar mass black holes to form more efficiently than in the field.
Natal kicks on formation may lead to black hole ejection, but this is likely a lesser
effect than with neutron stars, and some pulsars are evidently retained in clusters,
probably $O(10\%)$ or so (cf Sigurdsson \& Phinney 1995). 

It is interesting to note that there may in principle be a correlation between radio
loudness of pulsars and the natal kick magnitude, with neutron stars born with a low impulse 
being intrinsically radio quiet, and hence not observed in radio surveys of young neutron
stars. High resolution, deep x-rays surveys could constrain the existence of such a population
in the thin disk of the Milky Way.

Black holes provide for strong
dynamical evolution of the cores of globular clusters, and may drive some of the structural
anomalies we observe in clusters through relaxation and core collapse 
(cf Rasio this meeting). At late times, as the most massive cluster component
and the expected high number densities of BH in cluster cores, the interaction of black
holes with stellar binaries leads to exchanges, hardening and ejection, possibly leading
to severe depletion of the black hole population 
(Sigurdsson \& Hernquist 1993, Portegies Zwart \& McMillan 2000).
Some black holes may be retained through random chance; late segregation after the ejection phase is
over (or due to ejection to the cluster halo and long return times); or simply because one black hole
was significantly more massive initially and relatively invulnerable to ejection through dynamical
recoil by virtue of conservation of momentum (cf Miller \& Hamilton 2002).
The net effect is that some fraction of, or possibly all, the black holes 
in a typical cluster are ejected,
but any that remain tend to be massive, central and a member of a binary.

One constraint on the presence of black holes in globulars is the absence of a candidate
x-ray binary among the cluster binaries, though black hole binary candidates have now
been observed in extra-galactic globulars, so this is likely a small number statistic issue
(White, Sarazin \& Kulkarni 2002, Sigurdsson \& Hernquist 1993). We note that in those clusters in
which more than one black hole remains, we expect strong core heating by the black hole
binary interactions (cf Hemsendorf 2002), so multiple black holes are 
more likely in low density flat core clusters
(the compact remnant population may be in a high density cuspy sub-core, even if the optical
core is very flat and lower density).
If a ``most massive'' black hole is present, then it tends to grow significantly in mass,
and may be observed in the current epoch as an intermediate mass ($M_{BH} \gtorder 100 \msun$) black
hole, even if started off as merely a high end stellar mass black hole.

{\bf 2.1 BH in NGC~6752}\\
The case for a black hole in NGC~6752 is somewhat circumstantial and reviewed in by Possenti in
these proceedings (D'Amico et al 2002, Colpi, Possenti, \& Gualandris 2002). 
As a complement to the modeling done by that group, I have explored the
dynamical evolution of a pulsar binary ejected from the core of NGC~6752 due to an interaction
with a central black hole, as postulated by Colpi et al (2002). I have also explored structural
multi-mass King models for NGC~6752 with and without black holes, and modeled the interaction
history of single and binary black holes in model clusters with structural parameters
approximating those of NGC~6752. The basic model is as described in Sigurdsson \& Phinney (1995),
with 10 different (evolved) mass groups in mass-segregated equilibrium, 
and binaries evolved stochastically
in a fixed background cluster potentials, with Fokker-Planck diffusion calculated explicitly.
During the orbit of a binary an ``encounter probability'' is continuously calculated,
$R_{enc} = \sum_{\alpha} \int n_{\alpha} (r) \sigma( v_{bin}, v_* ) \vert {\bf v_{bin} - v_*}\vert f_{\alpha} (v_*)d^3 v_* $. I simulate the medium term evolution (on time scales shorter than
the cluster relaxation time) of an ensemble of binaries ($\sim 10^{2-3}$) following break-up,
hardening, dynamical recoil, exchanges and collisions to track the fate of binaries in a realistic
cluster model in some averaged sense. The current generation of models includes the possibility
of a population of single black holes, or the tracking of black hole binaries interacting with
the stellar population, or other black holes. 
\begin{figure}
\plottwo{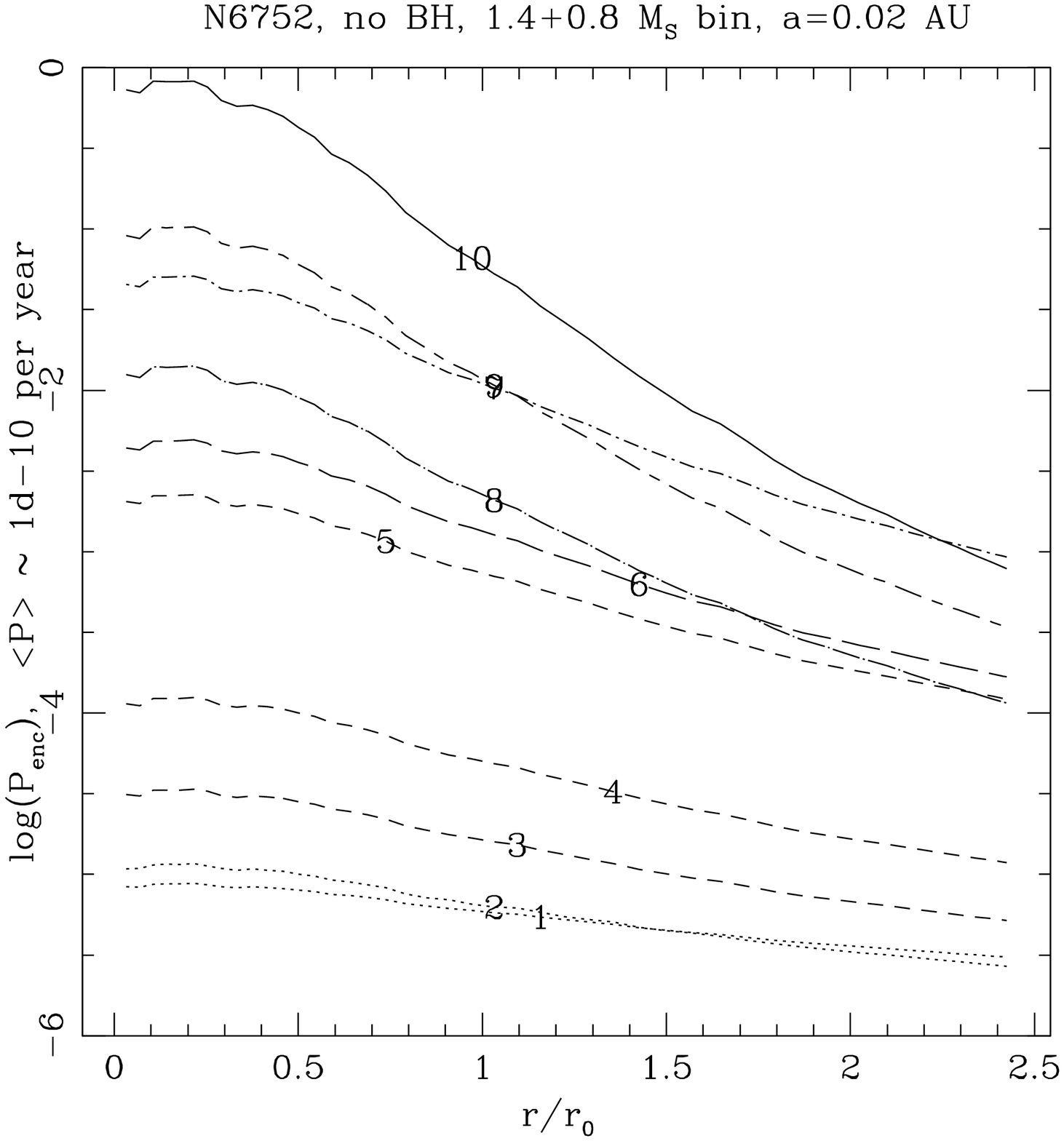}{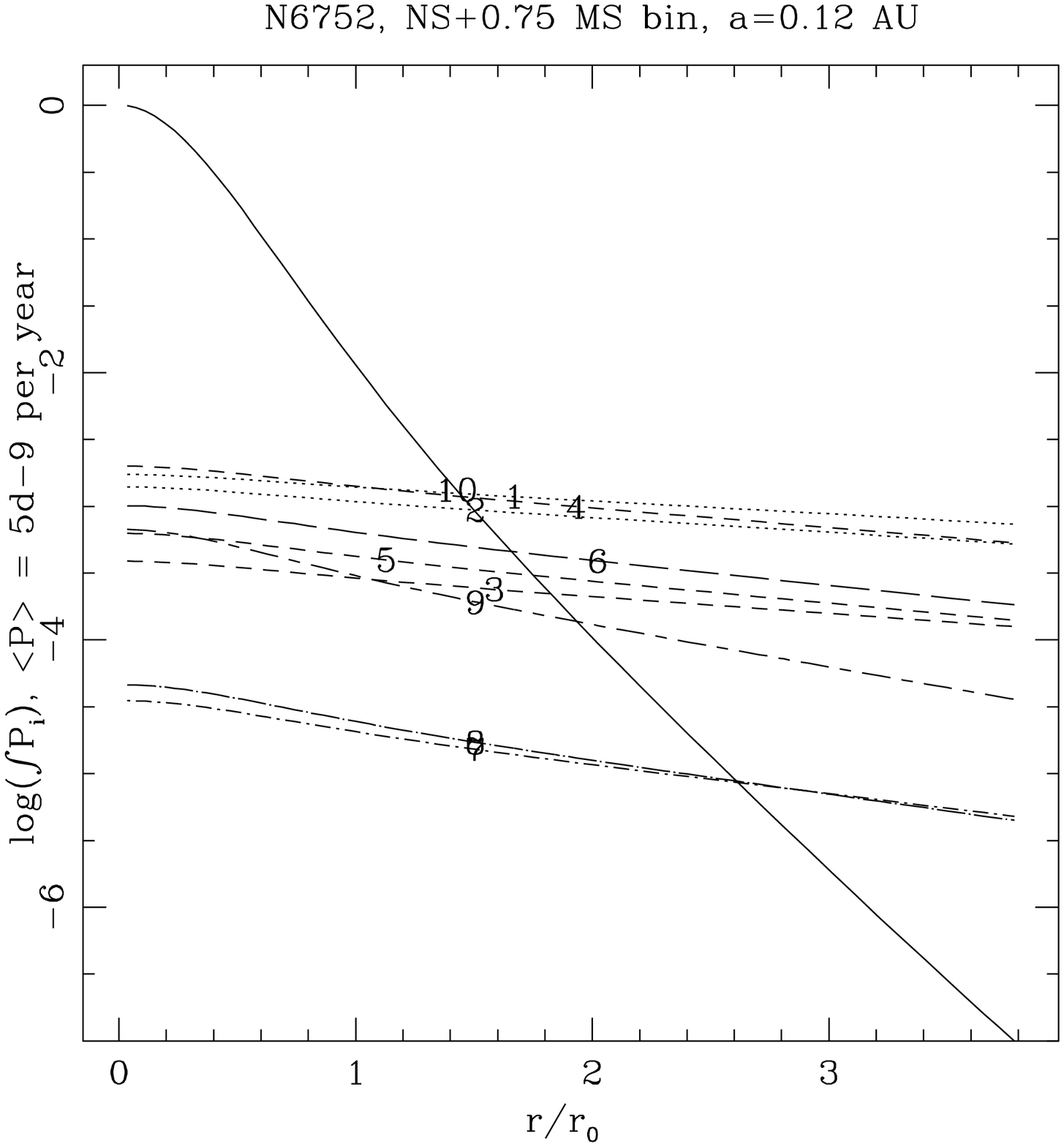}
\caption{Encounter probability for different mass groups for a pulsar binary
traversing a model of NGC~6752. If there is a(several) black hole(s) in the core
of the cluster, the integrated encounter probability is dominated by the black hole(s).
}
\end{figure}
Putting in a pulsar binary by hand, and forcing it to harden with associated dynamical
recoil of $\sim 6$ times the stellar dispersion, does indeed put it on a radial trajectory
persisting for $O(10^9)$ years, consistent with the observed projected position 
of PSR NGC~6752A in the cluster. If the companion of the pulsar is a turnoff mass main-sequence
star the binary sinks relatively rapidly back to the core through dynamical friction,
but a low mass companions allows for a persistent radial orbit. A key question is whether the
current companion was at its current mass at the time of ejection, or if it was a near turnoff mass
star which has undergone substantial forced mass transfer leaving only the partially evolved core.

As noted by Colpi et al, the binding energy of the current binary is barely adequate to achieve
the inferred $6\sigma $ recoil; there are however several ways to circumvent this limit, which
partially weakens the argument for an interaction with a massive black hole 
as the necessary channel to
absorb the momentum of the recoil. As noted before, the companion could have had more mass at
the time of recoil, with the post-encounter binary pericenter being inside the envelope of 
the secondary, leading to large mass loss and prompt post-encounter evolution; alternatively,
the pulsar binary might have interacted with a black hole binary, rather than a single binary,
in which case more binding energy is available for kinetic recoil.  In either case, the need for
a black hole to interact with is not eliminated, except possibly for very small probability
extreme binary-binary encounters; rather the mass-constraint on the black hole encountered is 
loosened.
\begin{figure}
\plottwo{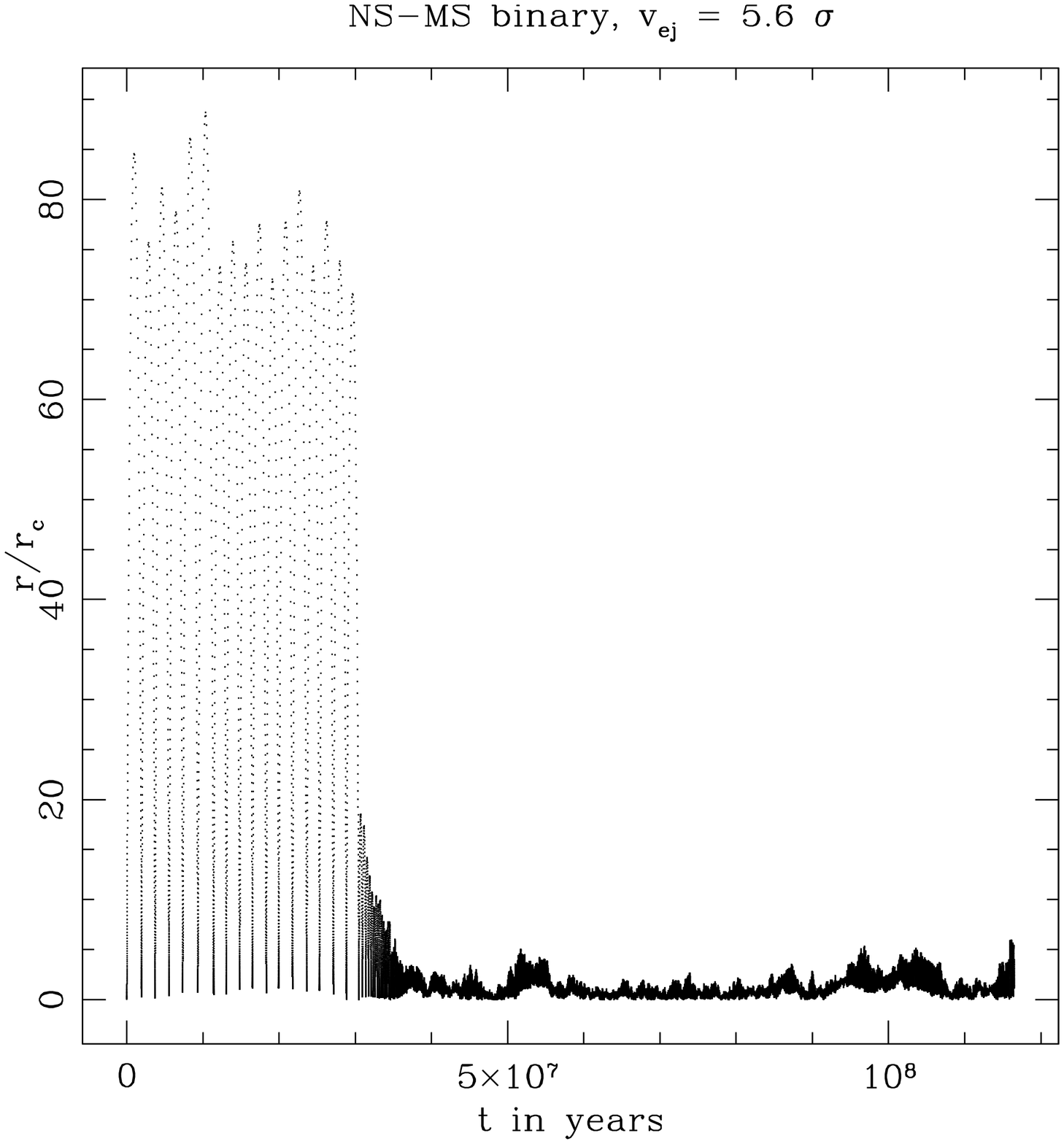}{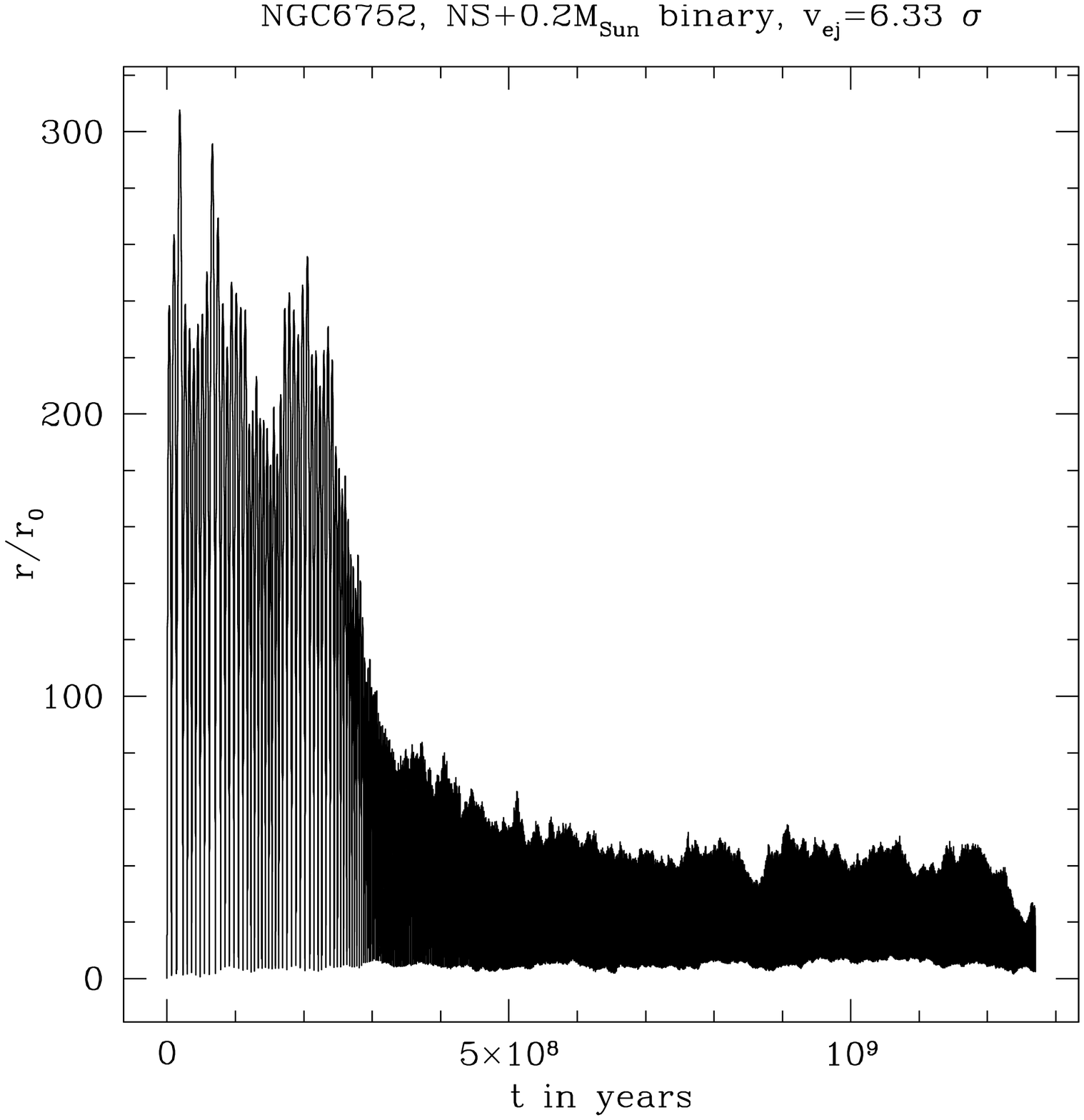}
\caption{Dynamical history of pulsar binary ejected from the core of NGC~6752.
The pulsar must have been ejected to more than 100 $r_0$, with initial speed over
$6 \sigma$. The pulsar companion must also have evolved rapidly ($\ll 10^8$ years) to low mass if it
was originally higher mass.
}
\end{figure}
Overall the existence of PSR NGC6752A is consistent with a formation history 
where it originated from a strong hardening and recoil event, most likely with
a moderately massive black hole or black hole binary in the core of NGC~6752.
I would argue that the pulsar companion was most likely originally a turnoff mass
main sequence star, possibly acquired in a previous exchange encounter, with the
current companion being the evolved remnant of the original companion, with evolution
probably forced due to rapid mass transfer in the binary post-interaction with the core black hole(s). 
The current position
of the pulsar binary is due to ejection from the core at about $6 \sigma $, with a dynamical lifetime
at its present position and mass of order $10^9$ years.

{\bf 2.2 BH binaries in GCs}\\
Three classes of black hole binaries may be present in globular clusters with some chance
of being observed.  A ``most massive'' black hole ($M_{BH} \gtorder 50-100 \msun$) formed
either by hierarchical growth from the most massive low mass black hole, or formed directly
at intermediate mass, may be present, and would efficiently exchange into any stellar binary
that made its way to the cluster core through dynamical friction. Dynamical recoil would
be low, and hardening would be rapid leading to short lifetimes. Pulsars could exchange
into such a binary if present initially as, eg. PSR-WD binaries with semi-major axis $\sim 0.1$ AU.
The resulting BH-PSR system would have $a \ltorder 0.2 AU$ and 
high ($\gtorder 0.9-0.99$) eccentricities. Dynamical evolution would be dominated by hardening
and lifetimes $\ltorder 10^7$ years are likely in dense clusters.

In low density clusters, a more modest ($M_{BH} \sim 10 \msun$) black hole may exist as
the ``lucky last'' black hole, which simply failed to eject during the black hole interaction phase.
Such a black hole would sink to the cluster core, and interact with any PSR-WD binaries, preferentially
those with $a \sim 1 $AU or so. The resulting BH-PSR system would form with relatively wide initial
semi-major axis and moderately high $\sim 0.9$ eccentricities. Dynamical recoil on formation would
be significant, enough to generally remove the system from the core, and 
it might spend few $\times 10^8$
years outside the core before returning through dynamical friction.

Finally, in core collapsed clusters we may see new black holes form through NS-NS mergers,
to form $2-3 \msun$ black holes, which may then exchange with other PSR binaries to form
a low mass BH-PSR binary, possibly somewhat displaced from the core through recoil for
timescales of order $10^8$ years. 

{\bf 3. Where might we find them}\\
In conclusion: we expect BH-PSR binaries to exist and I list 5 plausible candidate categories,
types I and II in the field, and 3 different types in clusters.
\begin{itemize}
\item{I} Type I field binaries are short lived and may not be generated efficiently at the present epoch.
\item{II} Type II field binaries (BH-MSP), ought to exist and should be seen within the next decade of
pulsar searches.
\item{GC-I} IMBH-MSP binaries may exist in core of massive clusters, formed by exchanges. 
They may be relatively short lived and
hard to observe.
\item{GC-II} BH-MSP binaries ought to exist in some fraction of intermediate density clusters, 
also formed through exchange,
possibly significantly displaced outside the core. {\it This is the ``best bet'' scenario for
observing BH-PSR binaries.}
\item{GC-III} New low mass BH may form in very dense GCs and form long lived BH-MSP systems
through exchanges, possibly spending some fraction of their lifetime just outside the core. 
\end{itemize}
%

I bet we will find a BH-PSR binary before Feb 7th 2008.

\acknowledgments
This research was supported in part by NSF grant PHY-0203046
and the Center for Gravitational Wave Physics, an NSF funded Physics Frontier Center
at Penn State.

\end{document}